\magnification=1200

\pageno=1
\centerline {\bf $p$-BRANES AS COMPOSITE ANTISYMMETRIC  }
\centerline {\bf TENSOR FIELD THEORIES     }
\medskip
\centerline {\bf Carlos Castro }
\centerline {\bf Center for Particle Theory, Physics Dept.   }
\centerline {\bf University of Texas}
\centerline {\bf Austin , Texas 78712}
\centerline {\bf World Laboratory, Lausanne, Switzerland}
\smallskip

\centerline {\bf March,1996 }
\smallskip
\centerline {\bf ABSTRACT}

$p'$-brane solutions to rank $p+1$ composite antisymmetric tensor field theories of the kind developed by Guendelman, Nissimov and Pacheva are found when the dimensionality of spacetime is $D=(p+1)+(p'+1)$. These field theories posses an infinite
dimensional group of global Noether symmetries, that of volume-preserving diffeomorphisms of the target space of the scalar primitive field  constituents.  Crucial in the
 construction of $p'$ brane solutions are the duality transformations of the fields and the local gauge field theory formulation of extended objects given by Aurilia, Spallucci and Smailagic. 
Field equations are rotated into Bianchi identities after the duality transformation is  performed and the Clebsch potentials associated with the Hamilton-Jacobi formulation of the $p'$ brane can be identified with the $duals$ of the original scalar primitive constituents. Different types of Kalb-Ramond actions are discussed and a particular  covariant action is presented which bears a direct relation to the light-cone gauge $p$-brane action. A simple derivation of $S$ and $T$ duality is also given. \medskip

\centerline {\bf I. INTRODUCTION}

\medskip

It has become clear in the past year that perturbative string theory will not answer some of the more vexing questions that it was set to answer. Why we live in four dimensions; which is the correct vacuum ; why is the cosmological constant zero after supersymmetry breaking; what is the underlying fundamental principle behind the string.....Nonperturbative string theory has captured considerable attention recently.  In particular, strings by themselves are not the only entities necessary to solve these questions  but, instead, 
 one  requires
to include all extended objects into the picture. Per example, in $D=10$ the 
string and the five-brane appear to be dual formulations of an unknown underlying physical theory. For a review on string solitons and duality properties in extended objects  one may see [1]. 

To this date we do not know what string theory is. A conjecture based on the theory of Scale Relativity was 
put forward in [2]. Such theory is based on the fundamental postulate that the Planck length is the absolute minimum, impassable, scale in nature invariant under dilations [3]; in the same way that Special Relativity is based on the speed of light being nature's  maximum velocity and an observer- invariant quantity. The principle of equivalence led to the General Theory of Relativity and, thus, to gravity. A new principle of equivalence based on the General Theory of Scale Relativity, where both dynamics as well as scales are incorporated into the picture, might lead to the underlying foundations in which to formulate string theory.

The aim in the present work is to show that $p$ branes are tightly connected with composite antisymmetric tensor field theories of the volume-preserving diffeomorphism group. Guendelman, Nissimov and Pacheva, GNP [4]
presented a new form of Quantum Electrodynamics in which  the photons are composites made out of scalar primitive constituents and where the role of local gauge symmetry was traded over to an infinite-dimensional $ global$ Noether symmetry : the group of volume-preserving diffeomorphisms of the target space of the scalar primitive constituents. 
Systems with infinite number of conservation laws allow to extract non-perturbative information and in some cases to solve the model exactly. The study of the Ward identities for infinite-dimensional global Noether symmetries to obtain non-perturbative information in mini-QED models has been analysed in [4, ANP].

In this letter we shall carry out the GNP  construction for higher rank $p+1$ antisymmetric theories and find that there are $p'$ brane solutions to these theories  in $D$ spacetime dimensions when $D=(p+1)+(p'+1)$. Crucial for the existence of these solutions is the duality transformation of the original fields and the  use of the gauge field  theory formulation of extended objects developed by Aurilia, Spallucci and Smailagic AS, ASS [5,6] based on original work by Nambu, Kastrup, Rinke, [7,8,9].

After having found $p'$-brane solutions we discuss the light-cone gauge. In the past [10] the light-cone gauge for $p$-brane actions   was conjectured to be related to a $new$ type of antisymmetric tensor gauge theories with an infinite-dimensional gauge group : that of $p$ volume-preserving diffs. 
Furthermore, these theories were $not$ of the Yang-Mills type ( except for the membrane). A covariant action is presented which bears a direct relationship to the light-cone $p$-brane actions [10] in the sense that it admits $p$ brane solutions ; i.e it furnishes a composite-antisymmetric tensor field theory formulation of extended objects. 

Extended solutions to local field theories have been studied in the past. What is $new$ in these $p'$ brane solutions is the fact that these are duals of a $composite$ antisymmetric tensor field made out of primitive scalar constituents. Matter fields can also be introduced. The notion of composite field theories might sound extraneous specially when one imagines  a photon as being a composite entity. Recently it was announced in Fermilab the possibility that the quark may be composite.
Hence, one should be open to new possibilities. The origin of duality is unknown in the theory of extended objects. We hope to  provide a step in solving this question.

Finally, in the last section,  we also  discuss some other topics as well, in particular, a simple derivation of the analog of the  $S,T$ duality symmetry in string theory  associated with the $p,p'$ brane solutions  and  the need to develop the formalism of antisymmetric tensor gauge theories over higher dimensional loop spaces to build the geometrical foundations of the $p'$ brane solutions presented here.

\smallskip

\centerline {\bf II. The Composite Antisymmetric Field  Theory }
\smallskip

Let us follow closely the field theory models in [4]. Consider a set of $p+1$ zero dimensional scalar fields $\phi^1(x),\phi^2(x)......\phi^{p+1}(x)$ on flat spacetime taking
values in a smooth manifold ${\cal T}^s$ with $s=p+1$. The group of volume-preserving
diffeomorphisms, $Diff_0 ({\cal T}^s)$ is defined such that 

$$\phi^a \rightarrow G^a (\phi) ; \epsilon_{b_1b_2....b_s}{\partial G^{b_1}\over \partial \phi^{a_1}}.......{\partial G^{b_s}\over \partial \phi^{a_s}} =
\epsilon_{a_1.......a_s} .\eqno (1)$$

Accordingly, the Lie algebra ${\cal D}iff_0 ({\cal T}^s)$ of infinitesimal volume-preserving diffs is
given by :

$$\{\Gamma^a (\phi);~ G^a (\phi) \sim \phi^a +\Gamma^a (\phi),~{\partial \Gamma^a \over \partial \phi^a} =0.\} \eqno (2)$$

where $\Gamma^a (\phi) $ is 

$$\Gamma^a (\phi) = {1\over (s-2)!}\epsilon^{abc_1.......c_{s-2}}{\partial \Gamma_{c_1.....c_{s-2}}\over \partial \phi^b}. \eqno (3)$$

When $s=2n$ the manifold ${\cal T}^{2n}$ may admit a symplectic structure $\omega_{ab}$ which can always
 be represented locally  by a canonical constant antisymmetric matrix. The associated (infinite-dimensional) Lie algebra of infinitesimal symplectic diffs of the infinite-dimensional   group of symplectic diffs ( diffs which preserve the symplectic structure ) is defined as :

$${\cal S}{\cal D}iff ({\cal T}^{2n}) \equiv \{\Gamma(\phi);~
[L_{\Gamma_1} , L_{\Gamma_2} ]= L_{\{\Gamma_1,\Gamma_2\}}\}. \eqno (4) $$
 
i.e. one has defined a Lie-Poisson structure : $[L_f,L_g ] =L_{\{f,g\}}$.

with :

$$\eta_1 =\Gamma^a_1{\partial\over \partial\phi^a}. ~\eta_2 =\Gamma^b_2 {\partial\over \partial\phi^b}. \eqno (5a) $$
where the componets , $\Gamma^a_1,\Gamma^b_2$, are given by the symplectic gradient :

$$\Gamma^a_1 =\omega^{ab}{\partial \Gamma_1 \over \partial \Phi^b}.~ \Gamma^b_2 =\omega^{ab}{\partial \Gamma_2 \over \partial \Phi^a}.\eqno (5b)$$
and  the Poisson bracket is :

$$\{\Gamma_1,\Gamma_2 \} \equiv \omega^{ab}{\partial \Gamma_1 \over \partial \phi^a}{\partial \Gamma_2
 \over \partial \phi^b}.\eqno (6)$$.  

$\omega^{ab}$ is the inverse of $\omega_{ab}$. 

The volume form of ${\cal T}^{2n}$ can be written as :
$\epsilon_{a_1......a_{2n}}= \omega_{[a_1a_2}.........\omega_{a_{2n-1}a_{2n}]}.$ Hence if the diffs are area-preserving they are automatically volume-preserving as well and one can define a Lie algebra of ${\cal D}iffs$  as before where eq-(5a,5b) are now  :

$$ \Gamma^a_1 = {1\over (s-2)!}\omega^{[ab}\omega^{c_1c_2}.......
\omega^{c_{s-1}c_{s-2}]}{\partial \Gamma_{1,c_1.....c_{s-2}}\over \partial \phi^b}. \eqno (7)$$

 The advantage of the Lie-Poisson structure  lies in the fact that one can quantize these theories by   
deforming the Poisson bracket via the Moyal bracket, per example. 

When $s$ is odd then the Lie algebra cannot be written in the Lie-Poisson  form but instead it can be  written in terms of the Lie derivative along a tangent vector $ v=v^at_a$ in ${\cal T}^s$  which maps $p$ forms into $p$ forms and tenors into tensors  :

$$l_v \equiv d~i_v +i_v~d.~i_v~\omega_{(p)} =pv^a\omega_{ab_2.....b_p}d\phi^{b_2}\wedge....
\wedge d\phi^{b_p}. \eqno (8) $$

Identifying $\eta_1 =\Gamma^a_1(\phi) t_a. ~\eta_2 =\Gamma^b_2 (\phi)t_b $ as two vector fields the Lie bracket of two Lie-derivatives   
is :

$$[l_{\Gamma^a_1t_a},l_{\Gamma^b_2 t_b}] =l_{[\Gamma^a_1t_a, \Gamma^b_2 t_b ]}. =l_{\Gamma^c_3 t_c}. \eqno (9) $$
with $\Gamma^a_1,\Gamma^b_2...$ given by eqs. like (3).
For further details about the gauging of  free differential algebras in connection to antisymmetric tensor gauge theories  see [11].

The canonical antisymmetric tensor ( volume form ) in  ${\cal T}^s$  defines 
an invariant  antisymmetric composite 
tensor field strength on spacetime satisfying the Bianchi identity, $dF=0,$
in terms of a potential :

$$A_{\mu_2......\mu_{p+1}}(\phi (x)) ={1\over (p+1)!}\epsilon_{a_1......a_{p+1}}\phi^{a_1}\partial_{\mu_2}\phi^{a_2}........\partial_{\mu_{p+1}}\phi^{a_{p+1}}.$$
$$F_{\mu_1.......\mu_{p+1}}(\phi (x)) =(p+1)! \partial_{\mu_1}A_{\mu_2.....\mu_{p+1}}(\phi).\eqno (10)$$

Antisymmetrization of indices is understood . The potential $A$ is defined modulo a total derivative of a $p-1$ composite-antisymmetric tensor $\Lambda _{\mu_2......\mu_{p+1}}(\phi(x))$. The field strength is invariant under finite field transformations belonging
to the  group   $Diff_0 ({\cal T}^s)$ as well as the addition of a total derivative  to the $A$ field :

$$ A_{\mu_2......\mu_{p+1}}(\phi) \rightarrow A_{\mu_2......\mu_{p+1}}(\phi)+ 
p!\partial_{\mu_2}\Lambda_{\mu_3......\mu_{p+1}}\eqno (11a)$$.

$$\Lambda_{\mu_3....\mu_{p+1}}=[(1-{1\over p+1}\phi^a
{\partial \over \partial \phi^a})\Gamma_{a_3...a_{p+1}} +{(p-2)!\over p+1}\phi^a {\partial \over \partial \phi^{a_3}}\Gamma_{aa_4...a_{p+1}} ]\partial_{\mu_3}\phi^{a_3}.....\partial_{\mu_{p+1}}\phi^{a_{p+1}}. \eqno (11b)$$ 

Let us
consider the  Lagrangian density ( over $D$-dim flat Minkowski spacetime ):

$$L=-{1\over e^2 (p+1)!}F^2_{\mu_1.......\mu_{p+1}}. \eqno (12)$$

This is the simplest Lagrangian invariant under global target 
 $Diff_0 ({\cal T}^s)$   
one could write describing a composite  antisymmetric tensor field theory made out of scalars $\phi^a$. One may set $e=1$ and we  should stress out that (12) does $not$ contain derivatives of time higher than two due to the antisymmetry of the indices [4]. The Lagrangian is a generalization of the usual non-linear $\sigma$ 
and WZNW models . A variation w.r.t $\phi^{a_1}$  yields the following eqs after premultiplying by a factor of $\partial_{\mu_{p+2}}\phi^{a_1}$ and using the Bianchi identity $dF=0$ :

$$F_{\mu_{p+2}\mu_2......\mu_{p+1}}\partial_{\mu_1}F^{\mu_1......\mu_{p+1}}=0.\eqno (13)$$

with 

$$F_{\mu_1.............\mu_{p+1}}\equiv \epsilon_{a_1......a_{p+1}}\partial_{\mu_1}\phi^{a_1}\partial_{\mu_2}\phi^{a_2}........\partial_{\mu_{p+1}}\phi^{a_{p+1}}.\eqno (14)$$

It is important to emphasize that despite the Maxwellian form of $F=dA$ the global symmetry, ${\cal D}iff_0({\cal T}^s)$, is $not$ abelian . The theory we are describing is $not$ of the Yang-Mills type; except for the membrane's case [10].

 We are going to find a special class of non-Maxwellian  solutions to eqs-(13) when the rank of the  $D\times (p+1)$ matrix $\partial_\mu \phi^a  \leq D-2$. In particular we are interested in $p'$ brane solutions of eqs- (13) for the case that $D=(p+1)+(p'+1)$. Due to the
rank condition the solutions of eqs-(13) are $not$ of the Maxwellian type 

$$\partial_{\mu_1} F^{\mu_1......\mu_{p+1}}=0 \rightarrow d^*F =^*J=0.
 \eqno (15)$$

This is reminiscent of the solutions to a homogeneous equation $M_{ij}X^i =0$.
Non-trivial solutions exist iff $det~M_{ij} =0$ otherwise $X^i=0$. 
$p'$-brane solutions are found first  by performing a $duality$ transformation :

$$ F^{\mu_1......\mu_{p+1}}={1\over (p'+1)!} \epsilon^{\mu_1......\mu_{p+1}\nu_1....\nu_{p'+1}}
G_{\nu_1......\nu_{p'+1}}({\tilde \phi}(x))\eqno (16)$$
where the ${\tilde \phi}(x)$ are the dual  primitive  scalars spanning a ${\cal T}^{p'+1}$ smooth target manifold.  
Eqs- (13) are then re-expressed in terms of the $dual$ fields $G$ :

$$G^{\mu_1\nu_2......\nu_{p'+1}}\partial_{\mu_1} G_{\nu_2......\nu_{p'+2}}=0
\eqno (17)$$

where 

$$\mu_{p+2}\rightarrow \nu_1.~\mu_{p+3}\rightarrow \nu_2......
\mu_D \rightarrow \nu_{p'+1}. \eqno (18)$$
$p'$-brane solutions to eq-(17) exist . They are based on  solutions to  a $local$  gauge field theory reformulation of extended objects given by ASS [5,6].  
The starting action is :

$$S=-g^2\int~d^Dx \sqrt {-{1\over (p'+1)!}W_{\nu_1...\nu_{p'+1}}W^{\nu_1......\nu_{p'+1}}}+{1\over (p'+1)!}W^{\nu_1...\nu_{p'+1}}\partial_{\nu_1}B_{\nu_2.....\nu_{p'+1}}. \eqno (19)$$

The above action is just the generalization to $p'$ branes of the action for a relativistic point particle coupled to an electromagnetic field :

$$S =\int~ds~+\int~ds~J^\mu A_\mu.~ \int~ds~J^\mu A_\mu\sim \int~dx^\mu A_\mu. \eqno (20)$$

The $W_{\nu_1....}$ is a totally antisymmetric tensor of rank $p'+1$ encoding the geometrical properties of the $p'$ brane and $g$ is a dimensional constant. 
The field strength is defined in terms of $B$ as $F=dB$. The physical dimensions are assigned :
$[W_{\nu_1...}] =[F_{\nu_1....}] =length^{-2}.~[g^2]=length^{2-D}$. 
$B$ field is a Lagrange multiplier enforcing a transversality constraint
( current conservation ) 
when one varies w.r.t $B$ : $\partial_{\mu_1} W^{\nu_1......\nu_{p'+1}}=0$. On shell the Lagrange multiplier $B$  becomes the $p'$ brane's gauge potential. A variation w.r.t $W_{\nu_1....}$  yields :

$$ g^2 { W_{\nu_1....\nu_{p'+1}}\over \sqrt{ -{1\over (p'+1)!}
W_{\nu_1......\nu_{p'+1}}W^{\nu_1......\nu_{p'+1}}}} +F_{\nu_1.....\nu_{p'+1}} =0. \eqno (21)$$                                     
The Bianchi identity for $B, d^2B=dF=0$, becomes, on-shell, the $p'$ brane eqs of motion iff one 
sets for a special solution to (21) the distribution-valued quantity :

$$
W_{\nu_1.....\nu_{p'+1}} = \kappa \int~d^{p'+1}\sigma ~
\delta^D (x -X(\sigma)){\vec X}_{\nu_1......\nu_{p'+1}}. \eqno (22)$$

where $\kappa$ is a dimensional constant and :

$${\vec X}_{\nu_1......\nu_{p'+1}} =\partial_{\sigma^1}X_{\nu_1}\wedge ......
\wedge \partial_{\sigma^{p'+1}}X_{\nu_{p'+1}}.
\eqno (23)$$
is the $p'+1$ tangent vector to the world-tube of the $p'$ brane.

Upon plugging (22, 23) into  (19) one recovers the Nambu-Goto-Dirac type of actions after the space-time integration  :

$$S(X)=-\kappa g^2~\int~d^{p'+1}\sigma~\sqrt{ -{1\over (p'+1)!}{\vec X}_{\nu_1......\nu_{p'+1}} {\vec X}^{\nu_1......\nu_{p'+1}}}.             \eqno (24)$$
with the effective surface tension $\rho =\kappa g^2$ which fixes the dimensions for $[\kappa] =l^{p-1}$. The square root expression is just the world-volume measure  as a result of the $p'$ brane's  embedding in spacetime. 
Using the Bianchi identity $dF=0$ in (21) and  taking  the pull-back 
onto the world tube of the $p'$ brane one recovers the eqs. of motion of the $p'$ brane :

$${\vec X}^{\nu_1......\nu_{p'+1}}\partial_{\nu_1}\Pi_{\nu_2......\nu_{p'+2}} =0. 
\eqno (25)$$
where $\Pi_{\nu_2......\nu_{p'+2}}$ is the geodesic field of the $p'$ brane defined below (33) [5,6].
Duality is indeed manifest in the sense that the Bianchi identities on-shell
  are equivalent to 
the equations of motion of the $p'$ brane.

It is not difficult to see that eqs-(25) are also  equivalent to the eqs. of motion of the 
Polyakov-Hower-Tucker 
types of action of a $p'$ brane ( with a cosmological constant for $p'> 1$). This is important to verify
since later we shall discuss the light-cone gauge eqs. of motion and action 
[10]. 
It was conjectured by [10] that the theory of extended objects is a new type of nonabelian antisymmetric gauge theories which is $not$ of the Yang-Mills type for $p'\geq 3$ and whose
gauge group is the infinite dimensional group of $p'$ volume-preserving diffs. For the membrane it coincided with a Yang-Mills theory
of the area-preserving diffs group. For a spherical membrane it was isomorphic 
to a suitable basis-dependent limit of $SU(N\rightarrow\infty)$ [12]. The latter allowed the authors
to quantize the membrane exactly [13]  since it was recognized that the spherical membrane admits
a reformulation in terms of the continous Toda theory with a natural $W_{\infty}$ symmetry
algebra [14]. Furthermore, the connection between non-critical $W_{\infty}$ strings and
membranes was established in [15] where the critical dimensions for the ( super) membrane was ($D=11$) $D=27$. For this reason we shall verify that (25) are the same
as the eqs of motion of the Polyakov type of  actions. For simplicity, take the closed membrane example. Starting with :

$$\epsilon^{abc}\partial_a X^\mu \partial_b X^\nu \partial_c X^\rho \partial_\mu
 \Pi_{\nu\rho\delta}=
\epsilon^{abc}\partial_bX^\nu\partial_cX^\rho \partial_a \Pi_{\nu\rho\delta}                                                 \eqno (26)$$
equation (25) is 

$$\partial_a [\epsilon^{abc}\partial_bX^\nu\partial_cX^\rho  
             \Pi_{\nu\rho\delta}]=0                  \eqno (27)$$
and using the definition of $\Pi$ below (33) one has :

$$ \epsilon^{abc}\partial_b X^\nu \partial_c X^\rho \Pi_{\nu\rho\delta}=
   \epsilon^{abc}\partial_b X^\nu \partial_c X^\rho  \epsilon^{mnr}
 \partial_m X_\nu \partial_n X_\rho \partial_r X_\delta 
[-det|\gamma_{ij}|]^{-1/2}.      \eqno (28)$$

Using the embedding condition for the induced world-volume metric :

$$\partial_bX^\nu\partial_m X_\nu =\gamma_{bm}.~\partial_cX^\rho\partial_n X_\rho =
\gamma_{cn}. \eqno (29) $$ 

and inserting the unit matrix :$\gamma_{ar}\gamma^{ar}=I$ in  eq- (28) yields  :

$$   [-det|\gamma_{ij}|]^{-1/2}\epsilon^{abc}\epsilon^{mnr}\gamma_{bm}\gamma_{cn}\gamma_{ar} \gamma^{ar}\partial_r X_\delta  =   $$

$$ [-det|\gamma_{ij}|]^{1/2}\gamma^{ar}\partial_r X_\delta. \eqno (30)$$
hence, after simplifying one finally arrives at the membrane eqs. of motion for Polyakov types of actions after  plugging (30) into (27)  :

$$  \partial_a (\sqrt{ -|\gamma|}\gamma^{ar}\partial_r X_\delta) =0.~\gamma_{ab}=\partial_aX^\mu\partial_bX^\nu \eta_{\mu\nu}.~\gamma^{ab}\gamma_{ab}=I.      \eqno (31)$$

After this detour let us go back to the dual form of the original eqs- (13,17) with the purpose of finding $p'$ brane solutions. Setting the l.h.s of the original eqs- (13,17) equal to the pull back of the 
Bianchi equation  one has along $x=X(\sigma)$ :

$$G^{\mu_1\nu_2...\nu_{p'+1}}\partial_{\mu_1} G_{\nu_2...\nu_{p'+2}}(x=X)=
\epsilon^{a_1...a_{p'+1}}\partial_{a_1}X^{\nu_1}...
\partial_{a_{p'+1}}X^{\nu_{p'+1}}\partial_{\nu_1}\Pi_{\nu_2...\nu_{p'+2}}=0. \eqno (32)$$

The $p'$ brane solutions to the r.h.s of (32) are [6] :

$$\Pi_{\nu_2.....\nu_{p'+2}}(x=X(\sigma))=-\rho {{\vec X}_{\nu_2.....\nu_{p'+2}}\over 
\sqrt {-{1\over (p'+1)!} {\vec X}_{\nu_2.....\nu_{p'+2}}{\vec X}^{\nu_2.....\nu_{p'+2}}}}. \eqno (33)$$ 

with the $p'+1$ vector  $ {\vec X}_{\nu_2.....\nu_{p'+2}}$ is given by 
(23). 

Eq-(33) is valid when the spacetime point $x$ has support on the $p+1$ world tube of the
$p'$ brane; i.e $x=X(\sigma)$. The canonical extension of the  solutions of the Bianchi identities to the whole  of spacetime was given by [6] in terms of the slope-field, 
$\Phi_{\nu_2......\nu_{p'+2}}(x)$.  This field when
evaluated on the $p'$ brane coincided precisely  with the $p'+1$ tangent-vector of 
the world tube of the $p'$ brane. The fluid analogy is that of a  vortex whose velocity
field  inside  a fluid coincides with the fluid's velocity at the points tangent to the vortex lines. The slope field at points outside the support of the $p'$ brane admitted the
unique extension :

$$\Phi_{\nu_2.....\nu_{p'+1}\nu_1}(x)={\int d^{p'+1}\sigma~\delta^D (x-X(\sigma)){\vec X}_{\nu_2.....\nu_{p'+2}}\over 
\int~d^{p'+1}\sigma~\delta^D(x-X(\sigma))}. \eqno(34) $$

It is the ratio of two singular distributions  in such a fashion that it is well defined everywhere in spacetime. The slope field is just the analog of the fluid's  velocity field ( which is well 
defined outside the vortex lines ).  The $D$ dimensional delta function can be 
evaluated as : 

$$\delta^D (x-X(\sigma)) ={\delta^{p'+1} (\sigma -\sigma')\over
\sqrt { -{1\over (p'+1)!}{\vec X}_{\nu_2.....\nu_{p'+2}} {\vec X}^{\nu_2.....\nu_{p'+2}}}} \delta^{D-p'-1} ({\vec x} -{\vec X}). \eqno(35) $$

The second delta function is over the transverse directions of the $p'$ brane. 
 Eq-(35) needs  to be regularized. Per example, the authors  [5,6] used a Gaussian regulator ; i.e a $p'$ brane with a certain thickness. Alternatively one could evaluate (35) in the static gauge.    
The quantitity (33)
$$\Pi_{\nu_1.....\nu_{p'+1}}(\sigma)=-\rho {{\vec X}_{\nu_1.....\nu_{p'+1}}
\over \sqrt {-(1/(p'+1)!) {\vec X}_{\nu_1.....}{\vec X}^{\nu_1.....}}}.
 \eqno (36) $$
is the so-called conjugate volume-momentum of the $p'$ brane. Upon its extension to all of  spacetime it becomes the geodesic-field associated with the gauge field theory description of extended objects [5,6] :

$$\Pi_{\nu_1......\nu_{p'+1}}(x=X(\sigma)) \rightarrow 
\Pi_{\nu_1......\nu_{p'+1}}(x) =-\rho {\Phi_{\nu_1.........\nu_{p'+1}}(x)\over
\sqrt{-{1\over (p'+1)!} \Phi_{\nu_1.........\nu_{p'+1}} \Phi^{\nu_1.........\nu_{p'+1}}}}. \eqno (37) $$

So in order to  extend  equations (32)  to the whole of spacetime it is useful first 
to  multiply  the r.h.s of (32) by $(-\rho /\sqrt{-(1/(p'+1)!) {\vec X}_{\nu_2.....}{\vec X}^{\nu_2.....}})$ and to replace  the $p'+1$ tangent vector,
 ${\vec X}_{....}(\sigma)$, 
by the slope field evaluated at $x$:

$$G^{\mu_1\nu_2......\nu_{p'+1}}\partial_{\mu_1} G_{\nu_2......\nu_{p'+2}}({\tilde \phi}(x))
=\rho {\Phi^{\mu_1\nu_2....\nu_{p'+1}}(x) \over \sqrt{-{1\over (p'+1)!} \Phi_{\mu_1.....}\Phi^{\mu_1.....}}}\partial_{\mu_1} \Pi_{\nu_2.....\nu_{p'+2}} (x) =0. \eqno (38)$$
Therefore, we can deduce that a particular solution (38) requires 
$G_{\nu_2......\nu_{p'+2}}({\tilde \phi}  (x))=$
$\Pi_{\nu_2.....\nu_{p'+2}}(x)$; i.e the dual ( in $D=(p+1)+(p'+1)$ dimensions ) of the original rank $p+1$ composite-antisymmetric field strength is the geodesic field , $\Pi$,   of the (dual)
$p'$-brane solution. And viceversa. This is one of the main conclusions of this work.

The correspondence with the string solitons discussed in the literature [1] is
straightforward. A $p$ brane couples to a $p+1$ form whose field strength is a 
$p+2$ form. The dictionary is established by shifting $p\rightarrow p+1$ so the 
duality condition is now : $D = (p+2)+(p'+2)$. Hence in $D=10$ the five-brane $p=5$
is dual to a string, $p'=1$. A three-brane is self-dual $p=p'=3$; In $D=6$ a string is self dual;  etc....
We shall study these models further when we discuss actions  with gauge-invariant Kalb-Ramond 
field interactions.

To conclude, once a $p'$ brane solution to (32) is given ( evaluated in the
$p'+1$ world- tube associated with the $p'$ brane ) one can extend it  over the whole of spacetime (34,37,38), and  go back to the original eqs of motion  for the
composite antisymmetric tensor field theory (13,14), and  have, automatically, a  solution for the 
field strength  after using (16) . Of course, there are $other$  solutions to eqs (13) besides the $p'$ brane solutions discussed so far. GNP presented simple non-Maxwellian  solutions in  the case that the target manifold was two dimensional for $D=2+1$ spacetime dimensions [4]. These displayed gauge-invariant 
photon mass generation ( via topologically massive Chern-Simmons theory ) where the generated mass was arbitrary and continous. A  sort of new effects was also discussed. 

The duality 
property is essential in constructing these solutions. Physically, a $p'$ brane moving in a $D$ dimensional spacetime has $D-(p'+1)$ physical degrees of freedom; i.e.  transverse to the world-tube of
the $p'$ brane. These transverse directions are precisely those where 
( some of ) the components of the composite antisymmetric tensor field strength live  in. Therefore, the dynamics of the $p'$ brane is encoded in the dynamics of the former theory via the duality transformations. Field equations rotate into Bianchi identities and the global Noether symmetry acting on the space of scalar constituents  is traded for a local gauge symmetry acting on the $p'$ brane's gauge potential : $B$. This will be important when we discuss the light-cone gauge.

The final step in this calculation  requires solving explicitly for the scalar fields
$\phi^a (x)$ ( after one has performed the canonical extension to the whole of spacetime ) :  

$$ F^{\mu_1...\mu_{p+1}}(\phi(x))\equiv \epsilon_{a_1...a_{p+1}}
\partial^{\mu_1}\phi^{a_1}(x)...\partial^{\mu_{p+1}}\phi^{a_{p+1}}(x)=$$
$$ \epsilon^{\mu_1...\mu_{p+1}\nu_1....\nu_{p'+1}}
\Pi_{\nu_1...\nu_{p'+1}}(x).    \eqno (39)$$

the latter (geodesic field) can be expressed as a Jacobian [6,8] :

$$ \Pi_{\nu_1......\nu_{p'+1}}(x) ={\partial (S^1(x),S^2(x).....S^{p'+1}(x))\over \partial (x^{\nu_1},
........x^{\nu_{p'+1}})} =G_{\nu_1....\nu_{p'+1}}({\tilde \phi}(x)). \eqno (40) $$

where the fields $S^1,......S^{p'+1}$ are called the Clebsch potentials [8].

In the Self Dual case where $p=p';D=2(p+1)$ and $^*F =F$ the equations  are greatly simplified since now the primitive scalars are just the Clebsch potentials. The first term of eq-(40) is invariant under the local gauge transformations  of the $p'$ brane gauge potential, $B$ :

$$\delta W^{\nu_1....\nu_{p'+1}}=0.~\delta B_{\nu_2....\nu_{p'+1}} =
\partial_{[\nu_2}\Lambda_{\nu_3.....\nu_{p'+1}]} (x). ~\delta \Pi=0. \eqno (41)$$
since on-shell $dB=-\Pi$.
Whereas the $G_{\nu_1....}({\tilde \phi}(x))$ dual field strength  is invariant under global  $p'+1$ volume-preserving diffs acting on the dual target space of scalar primitives. Under a change of ${\tilde \phi}^m \rightarrow {\tilde \phi}^m +\Gamma^m
 ({\tilde \phi})$ the field $G_{\nu_1....}({\tilde \phi}(x))$ does not change 
beacuse the Jacobian of the transformation in the ${\tilde \phi}$ space  is one. The definition of $G$ is the same as (14) for $F$ with $p$ replaced by $p'$ and $\phi\rightarrow {\tilde \phi}$. 

Therefore, the Clebsch potentials associated with the Hamilton-Jacobi formulation of the $p'$ brane  can be interpreted as being the $dual$ scalars ${\tilde \phi}(x)$ of the original scalar
$\phi (x)$ constituents  which comprised the $p+1$  antisymmetric tensor field in $D=(p+1)+(p'+1)$ dimensions. And viceversa : the Clebsch potentials associated with the $p$ brane  can be interpreted as being the dual scalars $ \phi(x)$ of the original scalar
${\tilde \phi} (x)$ constituents  which comprised the $p'+1$ antisymmetric tensor field strength.

It is useful  to  recall that in the point particle case one requires one Clebsch potential, the action : $S(x)$;  the Hamilton-Jacobi equation is $(\partial_\mu S)^2 +m^2 =0$. The $p'$ brane extension is [6] :

$$-{1\over (p'+1)!}[{\partial (S^1(x),S^2(x).....S^{p'+1}(x))\over \partial (x^{\nu_1},
........x^{\nu_{p'+1}})}]^2 +\rho^2 =0. \eqno (42)$$
where $\rho=\kappa g^2$ is the tension of the $p'$ brane.

This is the equation that the Clebsch potentials must obey. Once a solution
for the $S^1(x)....$ potentials  is known,  the $p'$ brane's volume-conjugate momentum is given automatically by : 
$$\Pi_{\nu_1...
\nu_{p'+1}} (\sigma)=
{\partial (S^1(x),S^2(x).....S^{p'+1}(x))\over \partial (x^{\nu_1},
........x^{\nu_{p'+1}})} (x=X(\sigma)). \eqno (43) $$

iff the $X(\sigma)$ are indeed solutions to  the $p'$ brane eqs. of motion. This is in  essence the Hamilton-Jacobi formulation of extended objects [5,6].

It is convenient to find the  relationship between the two effective surface tensions of the
$p;p'$ brane solutions to the corresponding antisymmetric field theories [5,6]  :

$$\kappa_p g^2_p =T_p.~\kappa_{p'} g^2_{p'} =T_{p'}.~[T_p T_{p'}]=length^{-D}. \eqno (44)$$

since the dimensions of $[T_p] =l^{-p-1};[T_{p'}]=l^{-p'-1}$ and $D-2 =p+p'$. If one chooses for fundamental scale the Planck length, $\Lambda_P$, and say one fixes $\kappa_p =\kappa_{p'}=1$; then the couplings $g_p,g_{p'}$ are inversely
proportional; i.e. a weakly  coupled $p'$ brane solution has a strongly coupled dual $p$ brane and viceversa.
  
In the string soliton literature [1] there is a shift of one unit in all the formulae as mentioned earlier so eq-(44) is replaced by :

$$[T_p T_{p'}]=l^{-D+2}. \eqno (45)$$

Finally, let us evaluate the action for these  $p'$ brane solutions. The Lagrangian density $-F^2_{\mu_1...}(\phi(x))\sim  - G^2_{\nu_1....}({\tilde \phi}(x))$ which is proportional to $-\Pi^2_{\nu_1.....}$. On the $p'$ brane's support the latter equals the tension squared, $+\rho^2$, yielding  a positive constant integrand which  makes the action positive definite.
In the Yang-Mills instanton literature one learnt  that  the use of the 
Schwarz inequality, analog of the vector equation :$ ||{\vec A}||||{\vec B}||\geq ||{\vec A}.{\vec B}||$,  yields in ${\cal M}=S^4$ :

$$\int~d^4x~F^2 =[\int~d^4x~F^2]^{1/2}[\int~d^4x~(^*F)^2]^{1/2}\geq 
\int~d^4x~F^*F \sim |winding~number|. \eqno (46)$$

Instanton solutions furnish an absolute-value lower bound to the action. In our case studied here one has slightly different results. In the self dual case $^*F=F$, the topological density is identically zero because $D> p+1$ and the action is zero. Zero actions appear naturally in many Topological Quantum Field Theories. In the other case studied we've just shown how the action is just propotional to the tension squared.

\medskip

\centerline {\bf III. The Light-Cone Gauge and Other Topics }

The light-cone gauge action for the super $p$
brane has been constructed in [10]. The bosonic piece omitting the zero modes is :

$$S=\int~d\tau\int~d^p\sigma[\partial_\tau X^I +u^a\partial_aX^I]^2 -det(\partial_aX^I\partial_bX^I). \eqno (47)$$
it agrees with earlier results [12]. The $p$ brane's clock, $\tau\sim X^+$ and the quantity $u^a$ is required to obey 
$\partial_a u^a =0$ in order for (47) to be invariant under $p$-volume-preserving diffs. The indices $a,b$ run over the spatial directions of the $p$ brane. The $I,J...$ run over the ``transverse'' $D-2$ spacetime directions. The lightcone gauge does not remove all degrees of freedom like occurs in the string. The residual gauge symmetries for $p\geq 3$ are :

$$\delta X^I =(\partial_b\Lambda^{ab})\partial_a X^I.~\Lambda^{ab}\rightarrow 
\Lambda^{ab}+\partial_c\Lambda^{abc}. \eqno (48a)$$
with the $\Lambda$'s being antisymmetric w.r.t $a,b...$ indices. 
Therefore, the net number of degrees of freedom are :

$$(D-2)-[p(p-1)/2-(p-1)(p-2)/2]=D-p-1. \eqno (48b)$$
as expected of a $p$ brane moving in $D$ spacetime dimensions

Earlier in (26-31) we have shown how the eqs. of motion for the Polyakov types of actions [10] agreed exactly wih the pullback of the Bianchi identity of the geodesic field of the $p$ brane. The light-cone gauge eqs. of motion must also agree. These are :

$$-(\partial_\tau +u^a\partial_a)^2X^I +\partial_a (hh^{ab}\partial_b X^I)=0.
~h\equiv det (\partial_aX^I\partial_bX_I)= det(h_{ab}). \eqno (49)$$
The equation w.r.t the ``gauge field'' of the $p$ brane, $\omega^{ab}$, where
$u^a=\partial_b\omega^{ab}$, ( locally) is :

$$ (\partial_a (\partial_\tau +u^a\partial_a)X^I)\partial_b X^I -a \leftrightarrow  b=0. \eqno (50)$$ 
and the $X^-$ coordinate obeys :

$$\partial_aX^- \sim (\partial_\tau +u^a\partial_a)X^I\partial_aX_I.$$
$$(\partial_\tau +u^a\partial_a )X^- \sim [(\partial_\tau +u^a\partial_a)X^I]^2
+h. \eqno (51)$$

The conclusion in [10] was that the net number of physical degrees of freedom of a $p$ brane, $D-p-1$, was linked to those of an antisymmetric tensor gauge theory whose gauge group was infinite-dimensionsl : $p$-volume-preserving diffs. Furthermore, for $p\geq 3$, the theory was $not$ of the Yang-Mills type. This was conjectured $after$ the lightcone gauge was chosen. What is this new theory ?

Let us start with the following manifestly covariant action which exhibits both of the geometric/gauge features of extended objects as the candidate action :

$$-\lambda^2 \int~d^Dx~\sqrt {-{1\over (p+1)!}F_{\mu_1.....\mu_{p+1}}(\phi(x))F^{\mu_1.....\mu_{p+1}}(\phi(x))}. \eqno (52)$$

The action is invariant under $p+1$ gobal volume-preserving diffs on the  space of scalar primitives. It has not the Yang-Mills form and the associated Lie algebra, ${\cal D}iff_0$,  is not abelian.
A study reveals that the self-dual solutions, $^*F=F;p=p'$
are compatible with the $p$ brane solutions to (52). A variation w.r.t $\phi$'s
yields in this case :

$$F^{\mu_1.....}\partial_{\mu_1}{F_{\mu_2.....\mu_{p+2}}\over \sqrt {F_{\mu_2.....\mu_{p+2}}(\phi(x))F^{\mu_2.....\mu_{p+2}}(\phi(x))}}=0. \eqno (53)$$

A particular class of $p$ brane solutions to (53) is given when $F$ equals the $distribution$-valued quantity given in (22) :

$${F_{\mu_2.....\mu_{p+2}}\over \sqrt {F_{\mu_2.....\mu_{p+2}}(\phi(x))F^{\mu_2.....\mu_{p+2}}(\phi(x))}}\sim {W_{\mu_2.....\mu_{p+2}}\over \sqrt {W_{\mu_2.....\mu_{p+2}}(x)W^{\mu_2.....\mu_{p+2}}(x)}}. \eqno (54)$$
so after substituting $F=W$ into (52) one recovers the Nambu action after the spacetime integration is performed. The effective tension is $\kappa \lambda^2$. Solutions in terms of the duals $G_{\nu_1...}({\tilde \phi}(x))$ exist as well in the case that $F$ is not self-dual. One simply may follow the same  steps as those taken in the previous section. There are $other$ types of solutions to (53) besides these ones. The action (52) for the particular solution given in (54) is the $p+1$ world-volume of the $p$ brane embedded in spacetime. When the light cone gauge is chosen (as any other gauge ) the number of degrees of freedom is reduced. After setting the $p$ brane's clock to agree with the $X^+$ spacetime direction one has the residual freedom to embed the spatial $p$-volume in the remaining $D-1$ directions and the physical degrees of freedom are the same as before, $D-1-p$.   

The light-cone-gauge  eqs. of motion (49) are the same as eq-(31) once the light-cone gauge is taken and a suitable parametrization of the $p+1$ world-volume metric is chosen. One must not confuse the indices used in (31) with those in (49). The field $u^a$ in (49) originates from the world-volume metric's parametrization into a spatial plus temporal slice. Since we have shown that (31) is exactly the same as the pullback of the Bianchi identity, which is just eq-(53), then we can infer that the action (52) in the light-cone gauge
must  reproduce the same light-cone-gauge eqs. of motion as (49) for the special $class$ of solutions of (53) given in (54). To conclude, the covariant action (52) represents a covariant composite antisymmetric tensor field theory formulation of  the $p$ brane. It displays the desired features
sought in [10]. 

What was manifest in the light-cone gauge of the spherical  membrane [12] was that it had a $correspondence$, not an identity, with a 
$D-1~SU(\infty)$ Yang-Mills theory dimensionally reduced to one temporal dimension. This didn't include the zero modes. The spacetime coordinates acquired the role of gauge potentials where the field $\omega^{ab}$ played the role of the
$A_0$ Yang-Mills field. Since one had a $D-1$ Yang-Mills theory the physical degrees of freedom were then ,$D-1-2=D-3$, as expected from a membrane. It was in this fashion how the supermembrane was viewed as a Supersymmetric Gauge Quantum Mechanical Model whose gauge group was the area-preserving diffs. Conversely,we could fix from the start the light-cone gauge of (52) and find a special $class$ of light-cone gauge solutions (eq-(54)) that should correspond to the light-cone gauge eqs. of motion of a $p$ brane. This justifies the use of 
eqs-(26-31).        

Instead of starting with the original action $F^2_{\mu_1......}(\phi(x))$ one could simply begin  with  
a  composite-antisymmetric Kalb-Ramond field theory   written in terms of a set of $p+2$ primitive scalars $\varphi (x)$ as folows : 

$$S= -{ 1\over (p+2)!}\int~d^Dx~H^2_{\mu_1.....\mu_{p+2}}(\varphi (x))+$$
$$\int~d^{p+1}\sigma~{1\over (p+2)}\epsilon_{a_1.....a_{p+2}}\varphi^{a_1}(x(\sigma))
\partial_{\sigma^1}\varphi^{a_2}(x(\sigma))....\partial_{\sigma^{p+1}}\varphi^{a_{p+2}}(x(\sigma)). \eqno (55)$$

The first integral has the same form of an antisymmetric tensor field theory and the second integral is just of the form : 

$$\int~d\Sigma^{\mu_1.....\mu_{p+1}}{\cal A}_{\mu_1....\mu_{p+1}}=
\int~d\sigma^1\wedge......\wedge d\sigma^{p+1}{\cal A}_{\sigma^1.....\sigma^{p+1}}.   \eqno (56)$$
based on the fact that a $p$ brane ( through its $p+1$ world-volume) couples to a $p+1$ Kalb-Ramond form 
${\cal A}$ whose field strength, $H$, is a $p+2$ form.

An alternative action is to augment  the original action for the composite antisymmetric tensor field of rank $p+1$, $F^2_{\mu_1.....}(\phi(x))$,  by a $standard$  gauge invariant Kalb-Ramond interaction with  a $p'$ brane  :

$$\kappa \int~d^Dx~ J^{\nu_1.....\nu_{p'+1}}{\cal A}_{\nu_1....\nu_{p'+1}}(x)-
{1\over (p'+2)!}\int~d^Dx~H^2_{\nu_1.....\nu_{p'+2}}(x).\eqno (57)$$

where now  the Kalb-Ramond field is a  standard fundamental field  instead of a composite one. In this fashion one makes  contact with the standard results in the string soliton literature [1]. The eqs. of motion are an extension of 
(13,14,17).
Alternatively, one could start with the candidate action (52) and add eq-(55) with an  ordinary Kalb-Ramond field instead of a  composite one. Such an action will represent a $p$ brane interacting with a Kalb-Ramond field. These alternative actions and the solutions to their eqs. of motion is worth looking into.

It has been discussed in the string case  [5] that when Kalb-Ramond interactions are introduced, the string's geodesic field $no$ longer satisfies the Bianchi identity, $d\Pi=0$,  but instead it is proportional to the Kalb-Ramond field strength : $d\Pi =-\kappa H_{\mu\nu\rho}$ and after projecting the field strength  onto the string world sheet one obtains the analog of the Lorentz force equation. i.e. the geodesic field of the string can be $absorbed$ into the redefinition of the
Kalb-Ramond field : 
$\kappa {\tilde A}_{\mu\nu}=\kappa A_{\mu\nu} +\partial_\mu B_\nu $; so that the ${\tilde H}=H$. For this reason it  is warranted to study these alternative actions further. The fact that 
electric-magnetic duality 
requires that $dF=\omega$ instead $dF=0$ [1] is consistent with $d\Pi =-\kappa H$; i.e.  by reabsorbing the geodesic field into a redefinition of the Kalb-Ramond field one is being compatible  with the string-soliton picture results. The ``electric'', ``magnetic'' charges  :

$$e=\int_{\partial M}~^*F=\int_M~^*J.~~~g=\int_{\partial M}~F=\int_M~\omega. \eqno (58)$$ 
should be part of a larger set of conserved charges.
The electric charge is conserved by virtue of the field eqs. whereas the magnetic charge is conserved by virtue of a global topological  conservation law. 
GNP [4] have an infinite number of conserved currents due to the infinite-dim
global Noether symmetry associated with the infinite-dim group of volume-preserving diffs of the target space. It is not known whether this would be sufficient to solve the non-perturbative theory exactly.  

Are there other types of antisymmetric tensor field theories which display the requirements of [10] ? It has been pointed out [5] that it is impossible to introduce non-abelian charges to the string  geodesic field; i.e it was impossible to ``colour''it. The coupling of the string to Yang-Mills fields occurs only  at the worldlines of pointlike boundaries. The colour degrees of freedom disappear from the Nambu action and are completely reabsorbed in the definition of the string tension. Non-abelian antisymmetric tensor gauge theories have been introduced in [16,17] with both vector-like and tensor gauge symmetries. 
Despite the presence of a non-abelian vector gauge symmetry there is the  problem  that the tensor-gauge symmetry is $abelian$. Pressumably, a non-abelian tensorial  gauge symmetry could be built to match the new non-Yang-Mills theories
conjectured in [10]. As we have discussed earlier, 
the action in eq-(52) displays the  new features required in [10]. The clue has been  to trade the global  volume-preserving diffs. over a local gauge symmetry  by introducing the scalar primitive  constituents as the fundamental fields. The problem of gauging the  closed string was solved by the introduction of the heterotic string. In principle the idea of compactifying some of the closed string coordinates  in an internal self-dual Lorentzian lattice should allow to ``colour' the geodesic field.

Quantization of extended objects is notoriously difficult due to the intrinsic nonlinearities. In [13] we were able to quantize the spherical membrane exactly based on its equivalence to a continous $sl(\infty)$ Toda model and the natural action of a $W_{\infty}$ symmetry. In the string case, quantization has been based on studying the spectrum of elementary particles which is interpreted as the quantum fluctuations around a classical background solution. Clearly, this is not the most ideal situation because one would like a background-indepent formulation of string theory. AAS [20] have focused on the transitions between different string configurations instead; i.e. on the geometric/topological properties of the string manifold : a loop space. A loop space representation for the string propagator was presented [20] based on a Hamilton-Jacobi formulation of string dynamics [8]. Above it was discussed that Hamilton-Jacobi eq-(42) contains the dynamics of the extended object; in!
 the string case,  by introducing the quantum operators ;

$${\hat p}_\mu (s) \sim ih{\delta \over \delta x^u (s)}.~{\hat H}\sim ih{\partial \over \partial A}. \eqno (59)$$
where $s$ is the loop space parameter, $A$ the area of the loop, allowed [20] to write down the quantum string kernel. The generalization of [20] to higher dimensional extended objects would require to work with higher dimensional loop spaces.
Unfortunately the mathematics of these loop spaces is virtually unknown. As unknown as it is the exact quantization of $p$ branes and their exact non-perturbative spectrum. The fact that Scale Relativity sets a limit on the lower scale in Nature to be the Planck length, $\Lambda_P$,  suggests that the quantization of these loop spaces must be such that the sizes of these $p$-loops must be quanta of $\Lambda_P^{p+1}$. Early work on loop spaces was undertaken in [17,18,19].  

A less ambitious project is for now to study the properties of the composite antisymmetric tensor field theories discussed in the present work. Supersymmetry is incorporated by using superspace methods. The scalars $\phi$ become superfields. To the action (52) one adds the Wess-Zumino terms. Gravity is also incorporated by replacing ordinary derivatives by covariant ones and including the Einstein-Hilbert action. Open $p$ branes have been studied in [6]. The inclusion of matter by [4].   
Work along these directions is in progress. The study of solutions of the Hamilton-Jacobi equation (42) is perhaps the most important project
because it furnishes solutions to the dual fields ,${\tilde \phi}(x)$, and the original $\phi (x)$ constituents for the self-dual $p$ brane solutions.

\centerline {\bf Final Note : S and T duality}

For convenience purposes set the couplings : $e=e'=1$ which appear in the respective actions of the rank 
$p+1$ and $p'+1$ composite-antisymmetric field strengths : $F^2(\phi);G^2({\tilde \phi}).$ One can deduce from the 
Hamilton-Jacobi equation associated to the $p'$ brane solutions to the
$F^2$ Lagrangian equations of motion ( where $F$ is  the pullback to spacetime of the volume form in the $\phi$ space )  that :

$$||-F^2_{\mu_1...\mu_{p+1}}(\phi)||\sim 
||-G^2_{\nu_1...\nu_{p'+1}}({\tilde \phi})||\sim T^2_{p'}=
(\kappa_{p'}g^2_{p'})^2.\eqno (A.1)$$

Similarily, one learns from the  Hamilton-Jacobi equation for the $p$ brane solutions to the $G^2$ Lagrangian equations of motion ( where $G$ is  the pullback to spacetime of the volume form in the ${\tilde \phi}$ space )  that  ;

$$ ||-G^2_{\nu_1...\nu_{p'+1}}({\tilde \phi}')||\sim 
||-F^2_{\mu_1...\mu_{p+1}}( \phi ')||\sim T^2_{p}=
(\kappa_{p}g^2_{p})^2.\eqno (A.2)$$
$\phi (x);{\tilde \phi}(x)$ are the solution set for eq-(A.1) and
$\phi ' (x);{\tilde \phi}'(x)$ are the solution set for eq-(A.2) and should not be confused with the former solutions to eq-(A.1). From the above equations one can infer immediately due to the fact mentioned earlier :
$T_p T_{p'}=\Lambda^{-D}$ which is the statement of S-duality, strongly coupled value of $g_p$ corresponds to weakly coupled value of $g_{p'}$ and viceversa  for $\kappa_p =\kappa_{p'}=1$, that  :

$$||-G^2_{\nu_1...\nu_{p'+1}}({\tilde \phi}')||||-G^2_{\nu_1...\nu_{p'+1}}({\tilde \phi})|| \sim T^2_p T^2_{p'}=\Lambda^{-2D}. \eqno (A.3)$$

and similarily :

$$||-F^2_{\mu_1...\mu_{p+1}}(\phi)||||-F^2_{\mu_1...\mu_{p+1}}( \phi ')||\sim T^2_p T^2_{p'}=\Lambda^{-2D}. \eqno (A.4).$$
Taking the square-root to both sides of (A.3,A.4) yields  :

$$\Omega_{p'+1}({\tilde \phi}')\Omega_{p'+1}({\tilde \phi})\sim \Lambda^{-D}.
~\Omega_{p+1}(\phi')\Omega_{p+1}(\phi)\sim \Lambda^{-D}.\eqno (A.5)$$
If one wished to reintroduce  the couplings, $e,e'$, one just inserts factors of $e'^{-2},e^{-2}$ in (A.5) repectively.

Therefore, we can infer that the pullback to spacetime of the respective volumes in the $\phi$ and ${\tilde \phi}$ spaces given in terms of the solution-set
to the Hamilton-Jacobi equations associated with the $p,p'$ branes, eqs-(A.1,A.2), are $inversely$ related which is just another analog  of the $T$ duality 
symmetry in string theory :$R\leftrightarrow \alpha '/R$; where $\alpha '$ is the inverse string tension.
Hence, in this approach to extended objects we can see that $S$ and $T$
 duality are interconnected and are already manifest from the very beginning. 
One must emphasize that we are not asserting that $S,T$ duality are trivial symmetries. The solution-set to eqs-(39,40) from which  eqs-(A.1,A.2) are obtained, is far from trivial. These  equations are highly non-linear. Once a solution set to (39,40) is $found$ it automatically 
 obeys  eqs-(A.3,A.4,A.5) and, hence, it satisfies the $S,T$ duality conditions.

We cannot say anything yet about $U$ duality until we have the supersymmetric formulation and the compactification solutions from $D=11$ to lower dimensions. The fact that $S$ and $T$ duality are interconnected , as we have shown, should  be  a reflection of the so-called duality of dualities in the string-solitons literature which exchanges $S\leftrightarrow T$ in certain lower dimensional superstring vacuum solutions. The real test lies now in the quantization program in addition to solving explicitly for the Clebsch potentials; i.e. for the $\phi,{\tilde \phi}$ scalars. More precisely : Does the quantization preserve the 
$S,T$ duality conditions ?

\centerline {\bf Acknowledegements}

We are greatfuly indebted to G. Sudarshan, C. Ordo${\tilde n}$ez, J. Boedo, M. Bowers  for their advice
and support and to J. Pecina for his help
in preparing this manuscript. This work was possible thanks to an ICSC 
World Laboratory grant.

\centerline {\bf REFERENCES}

1. M.J. Duff, R.R. Khuri, J.X. Lu : Physics Reports {\bf 259} (1995) 213-326.

2. C. Castro : `` String Theory, Scale Relativity and the Generalized Uncertain

ty Principle ``; submitted to Physics Letters B. 

3. L. Nottale : `` The Fractal Structure and the Microphysics of Space Time :

Towards the Theory of Scale Relativity ``. World Scientific. 1992. 

Int. Jour. Mod. Physics {\bf A 7} no. 20 (1992) 4899.

4. E.I. Guendelman. E. Nissimov, S. Pacheva : `` Volume-Preserving Diffeomorphi

sms versus Local Gauge Symmetry `` hep-th/9505128. BGU-95/10/May-PH preprint.

H. Aratyn, E. Nissimov, S.Pacheva : Phys. Lett {\bf 255 B} (1991) 359.

5. A. Aurilia, A. Smailagic, E. Spallucci : Phys. Rev {\bf D 47} no. 6 (1993)

2536.

6. A. Aurilia, E. Spallucci : Class. Quan. Gravity {\bf 10} (1993) 1217.

7. Y. Nambu : Phys. Lett {\bf 102 B} (1981) 149.

8. H. Kastrup : Phys. Rep. {\bf 101} (1983) 1-167.

9.H. Kastrup, M. Rinke : Phys. Lett {\bf 105 B} (1981) 191. 

10. E. Bergshoeff, E. Sezgin, Y. Tanii, P.K Townsend : Annals of Phys. 

{\bf 199} (1990) 340.

11. L. Castellani, A. Perotto : `` Free Differential Algebras : Their Use in 

Field Theory and Dual Formulations `` hep-th/9509031. DFTT-52/95 preprint.

12. J. Hoppe : `` Quantum Theory of a Relativistic Surface `` MIT Ph.D thesis

1982.

13. C. Castro : `` An Exact Membrane Quantization from $W_{\infty}$ Symmetry ``

with the publishers of the  to Int. J. Mod. Physics A.

14. M.V. Saveliev : Theor. Math. Phys. {\bf 92} no. 3 (1992) 457.

15. C. Castro : `` $D=11$ Supermembrane Instantons, $W_{\infty}$ Strings and

the Super Toda Molecule ``. To appear in the J. Chaos, Solitons and Fractals.

April (1996).

16. D. Freedman, P.K. Townsend : Nucl. Phys. {\bf B 177} (1981) 282.

17. L. Smolin : Phys. Lett {\bf 137 B} (1984) 379.

18.P.G.O Freund, R.I. Nepomechie : Nucl. Phys. {\bf B 199} (1982) 482.

19.S. Deguchi, T. Nakajima : Prog. Theor. Phys. {\bf 94} no. 2 (1995) 305.

20.S. Ansoldi, A. Aurilia, E. Spallucci : ``String Propagator : a Loop Space 

Representation `` hep-th/9510133. UTS-DFT-95-4.

\end